# Measurement of human activity using velocity GPS data obtained from mobile phones

Yasuko Kawahata[1] Takayuki Mizuno[2] and Akira Ishii[3]

[1] Graduate School of Information Science and Technology, The University of Tokyo, Hongo, Tokyo 113-8654 Japan
[2] National Institute of Information, Hitotsubashi, Chiyoda-kuTokyo 101-8430, Japan

[3] Tottori University, Tottori 680-8552, Japan

`purplemukadesan@gmail.com`

**Abstract.** Human movement is used as an indicator of human activity in modern society. The velocity of moving humans is calculated based on position information obtained from mobile phones. The level of human activity, as recorded by velocity, varies throughout the day. Therefore, velocity can be used to identify the intervals of highest and lowest activity. More specifically, we obtained mobile-phone GPS data from the people around Shibuya station in Tokyo, which has the highest population density in Japan. From these data, we observe that velocity tends to consistently increase with the changes in social activities. For example, during the earthquake in Kumamoto Prefecture in April 2016, the activity on that day was much lower than usual. In this research, we focus on natural disasters such as earthquakes owing to their significant effects on human activities in developed countries like Japan. In the event of a natural disaster in another developed country, considering the change in human behavior at the time of the disaster (e.g., the 2016 Kumamoto Great Earthquake) from the viewpoint of velocity allows us to improve our planning for mitigation measures. Thus, we analyze the changes in human activity through velocity calculations in Shibuya, Tokyo, and compare times of disasters with normal times.

**Keywords:** people's moving speed, mobile phone, earthquake.

## 1  Introduction

Human activities in contemporary society can be recorded as digital data using smartphones and other electronic equipment. Therefore, it was possible to quantify the changing speed in daily activities of a large group of people based only on such data. Moreover, we can estimate how human activity will change in the event of a disaster. For example, Telepoint data can be considered as spatiotemporal data, including latitude and longitude information, on the development and decline of industries in various regions. Recently, it has become possible to know not only the posi-tion (latitude and longitude) of a person but also their speed and direction of movement from GPS position information obtained using a mobile phone. In recent years, research has begun on the movement speed of groups of people. Luis et al [1] and Bettencourt et al. [2] attempted to determine the relation between economic activity and humans' movement speed. Focusing on the relation between city population and walking speed, we considered walking speed as a factor of economic productivity and argued for a positive relation between walking speed and economic activity in any given area. In addition, Korsu et al [3] discussed urban planning from the viewpoint of moving speed.

In the present research, we focus on the effect of natural disasters such as earthquakes on human activity in Japan. We consider the changes in people's behavior at the time of a disaster (the 2016 Kumamoto Great Earthquake) from the viewpoint of speed and suggest the type of support that should be provided when disasters occur in other developed countries. As a case study, we also consider and analyze the changes in humans' velocity before and after a natural disaster in Shibuya, which is the central part of Tokyo and has the largest flowing population.

## 2    Velocity data

We obtain the mobile-phone location data from AGOOP Corporation in Japan. AGOOP was established in April 2009 and is owned by SoftBank Corp., which is a major mobile-phone service company. These location data include position information logs, both domestic and overseas, obtained with GPS. The GPS location information was acquired after obtaining written access permission from the users. These location data used here to generate two types of floating population data: 1. "point-type floating population data," which contain detailed GPS data as "moving points," and can be used to estimate the flow and trend of people; and 2. "mesh-type floating population data" that can extrapolate results from the application users to the total population of Japan for a certain period of time. These data can be utilized in various fields such as marketing of commercial facilities, economy and tourism policies, and disaster prevention and mitigation.

In this paper, we use the location data for Shibuya (downtown Tokyo) and Kumamoto, where a great earthquake took place in April 2016 [4] (Fig. 1). Tokyo is the capital of Japan and has a population of 13,686,371, reaching twenty million including neighboring big cities around Tokyo. On the contrary, Kumamoto is a prefecture located in the Kyushu area in the western part of Japan, and the prefecture's population is 1,765,940.

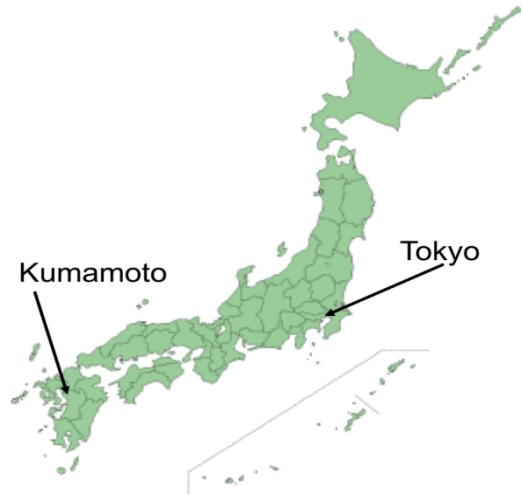

**Fig. 1.** Location of Tokyo and Kumamoto in Japan

## 3 People's daily velocity

First, we see the mean velocity of people every day. The difference in the daily mean velocity of the people in Shibuya, Tokyo, for October 2014 is shown in the Fig. 2. The mean daily value increases during the weekends, whereas weekdays show approximately the same low value. To be precise, the highest values are recorded on Fridays and Saturdays. In addition, in the middle of October, which includes three consecutive holidays, velocity is high from the day before the holidays, and the velocity on Sunday is similar to that on Monday.

Shibuya is one of the busiest downtown areas in Tokyo, which is a city where young people gather on holidays [5]. The area around Shibuya-ku is a cultural and market center (including Harajuku) gathering Japan's youth, and it had the highest levels of activity based on the data handled in this research. It has several scramble intersections that are great tourist spots and places where many people meet and interact on a daily basis. Therefore, activity is expected to be high on Fridays and Saturdays. As Sunday is followed by a weekday, it seems that activity will not be higher than Saturday.

**Fig. 2.** Daily mean velocity in Shibuya, Tokyo, for October 2014

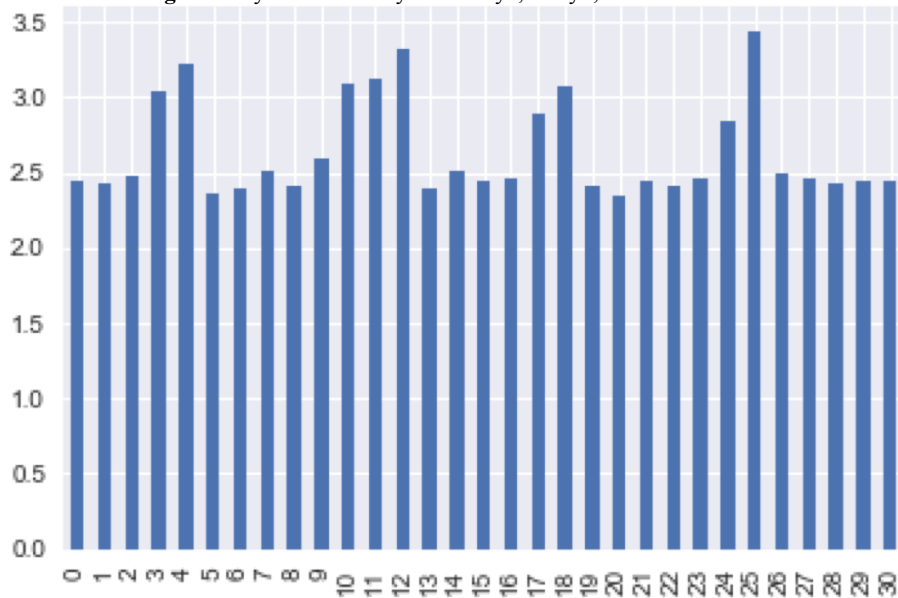

The following figure shows the mean velocity distribution for October 2015. Looking at this graph, few people are moving at high speed, i.e., using a car, a train, or a bus. Therefore, most people are walking in Shibuya. Therefore, the activity of people on Fridays and Saturdays shown in Fig. 2 represents the people walking in the downtown area of Shibuya.

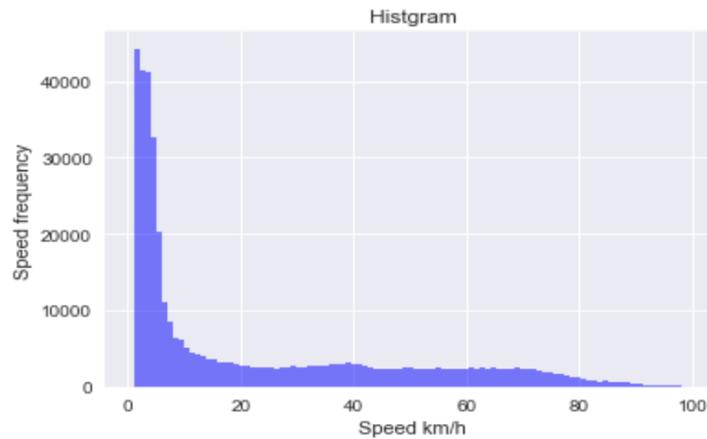

**Fig. 3.** Velocity distribution in Shibuya, Tokyo, for October 2015.

On the contrary, the people's mean velocity distribution in Kumamoto Prefecture is shown in the Fig. 4. This distribution is high, around 30–50 km/h, referring to car movement. Since slow speed refers to pedestrians, we infer that, in Kumamoto, people move on foot or use a car. In other words, in Kumamoto, which is a regional city, automobile transportation is extensive.

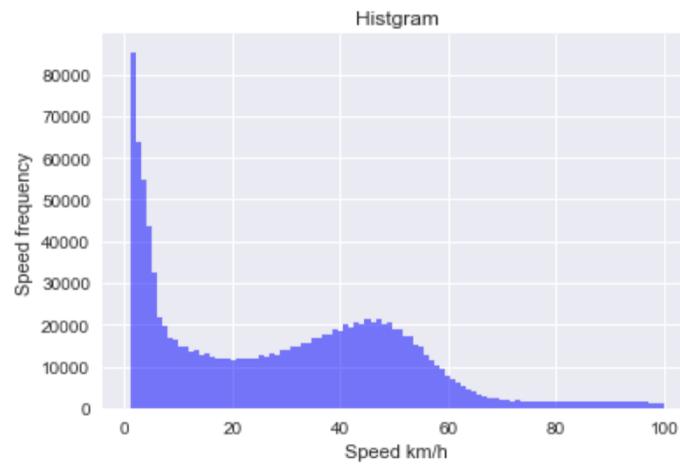

**Fig. 4.** Velocity distribution in Kumamoto Prefecture for April 2016.

Next, we study the difference in people's velocity throughout the day. Figs. 5 and 6 show the difference in the moving speed between Shibuya and Kumamoto in one day every hour. From the figure, it is understood that the traveling speed decreases during lunchtime and at night. This means that people do not move during lunch hours during the day and at night when they go home.

Therefore, we consider that the increase/decrease in velocity corresponds to an increase/decrease in human activity.

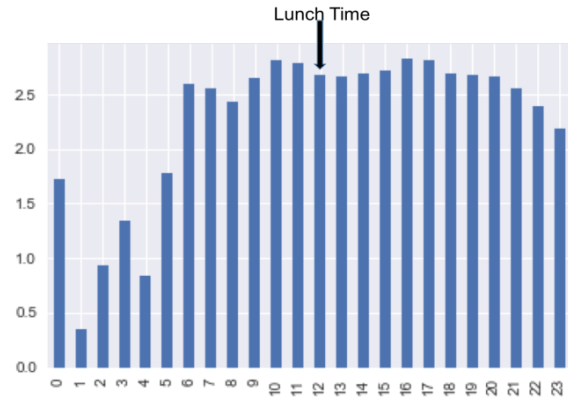

**Fig. 5.** People's moving speed of the day. Data for Tokyo Shibuya in October 2015.

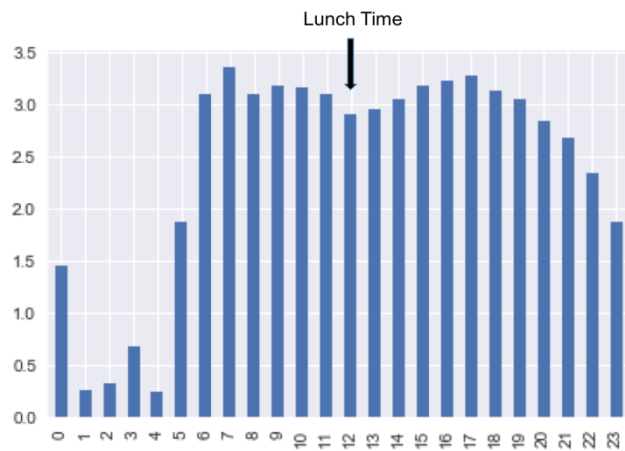

**Fig. 6.** People's moving speed of the day. Data for Kumamoto in April 2016.

## 4    People's moving speed on the day of the earthquake

To investigate people's activity with regards to the moving speed, we examine the day of the earthquake in April 2016. Kumamoto Prefecture in the western part and Oita prefecture in the east of Japan suffered big earthquakes on several days around April 15, 2016. The average travel speeds of people every day for April 2016 is shown in the Figs. 7 and 8. These graphs are drawn with data for Kumamoto Prefecture and Oita Prefecture, which was greatly damaged by the earthquake [4].

Indeed, people's velocity is low on April 15, when the major earthquake occurred. As the major earthquake caused great damage to both Kumamoto and Oita prefectures, a considerable amount of people's activities ceased almost completely immediately after the earthquake.

The low value of velocity for the day of the earthquake seen in these figures seems to suggest that the traveling speed data can be used as an indicator of human activity.

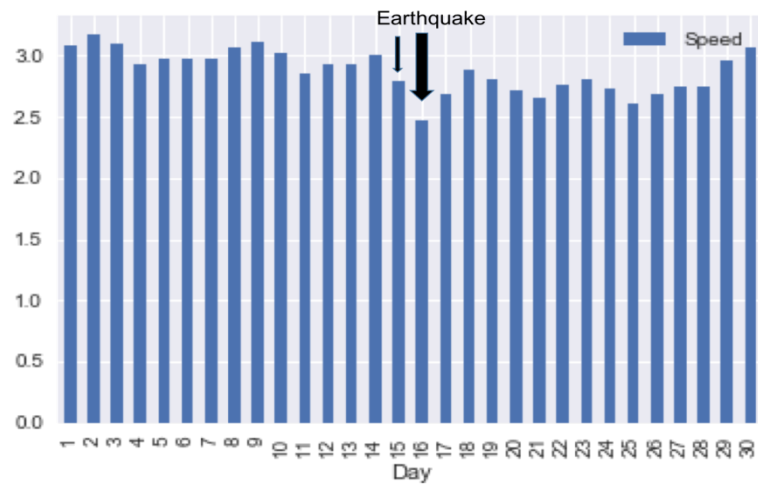

**Fig. 7.** Daily mean velocity in Kumamoto for April 2016. The arrows indicate the days of the earthquakes. The larger arrow indicates the major earthquake.

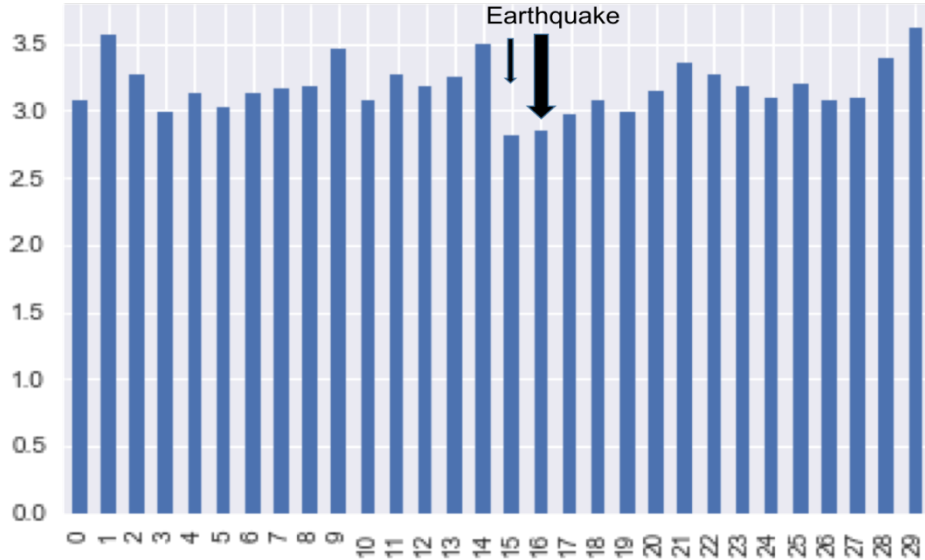

**Fig. 8.** Daily mean speed in Oita prefecture for April 2016. The arrows indicate the days of the earthquakes. The larger arrow indicates the major earthquake.

## 5    Discussion and Conclusion

In this paper, we focus on changes in the activities of people during emergencies using human movement information. To investigate whether people's activities occur at the moving speed, we investigate the day of the earthquake in April 2016. Indeed, human activity, as indicated by the daily mean velocity of the people in the two study areas, significantly decreases in the aftermath of a major natural disaster such as an earthquake. In particular, people and cars decreased their moving speed on the day of the earthquake. The lowering of the mean velocity results from the shift of the moving speed to a lower speed in the velocity distribution. This does not mean that people do not walk or that all cars run late. Instead, this decline in the mean velocity indicates that people stop their cars and take a break or that pedestrians stop walking and take a break.

Overall, we observe a decrease in mean velocity during lunchtime, at night, and on the day of the earthquake. People eat lunch and have a break during the day. People sleep at home at night. In addition, it is thought that the decrease in the mean velocity after the earthquake is caused by the difficulty in moving the car due to traffic infrastructure damages, and normal operations are stopped due to the damages caused by the earthquake.

Temporary decline in human flows and logistics results in a temporary decline in social and economic activities because, according to the input–output table [6, 7], all industries are related to logistics. It is thought that a decline in the average moving speed in a society is linked to a decline in economic activity. The fact that the average

9moving speed is decreasing during the daytime break, late at night, and the day of the earthquake seems to suggest that the average moving speed can be used as an indicator of the degree of socioeconomic activity.

Furthermore, from the results of Kumamoto and Oita, no significant speed reduction is observed during the weekend. This would mean that the weekend's economic activities are being activated in the form of consumer activities for many people. In addition, the weather does not significantly affect the decline in the average speed.

Examples of expectations for the observation of the decline in the average speed of society are festivals where the entire city are represented by Rio Carnival participants and at the time of a game of their own team during the World Cup.

In conclusion, we point out a temporary decline in human activity, as indicated by people's mean velocity, using mobile-phone location information. Loss of speed occurs when socioeconomic activities cease, such as during lunchtime, late at night, and in the event of a natural disaster such as an earthquake. Therefore, we propose using mean velocity as an indicator of economic activity.

## References

1. Bettencourt, L.M.A., Lobo, J., Helbing, D., Kühnert, C., West, G.B.: Growth, innovation, scaling, and the pace of life in cities. Proceedings of the National Academy of Sciences of the United States of America 104(17), 7301–7306 (2007).
2. Bettencourt, L.M.A., Lobo, J., West, G.B.: Why are large cities faster? Universal scaling and self-similarity in urban organization and dynamics. The European Physical Journal B 63, 285–293 (2008).
3. Korsu, E., Le Néchet, F.: Would fewer people drive to work in a city without excess commuting? Explorations in the Paris metropolitan area. Transportation Research Part A: Policy and Practice 95, 259–274 (2017).
4. 2016 Kumamoto earthquakes, https://en.wikipedia.org/wiki/2016_Kumamoto_earthquakes, last accessed 2017/05/16
5. Daniels, P.W., Ho, K.C., Hutton, T.A.: New economic spaces in Asian cities: from industrial restructuring to the cultural turn. Journal of Regional Science 53(2), 39–66 (2012).
6. Leontief, W.: Quantitative input and output relations in the economic system of the United States. The Review of Economics and Statistics 18, 105–125 (1936).
7. Ministry of Internal Affairs and Communications, Japan: 2011 Input-output table for Japan.